\documentclass[alpha-refs, serif]{wiley-article}

\usepackage[utf8]{inputenc}
\usepackage{siunitx}
\usepackage{float}
\usepackage{comment}
\usepackage{algpseudocode}
\usepackage[Algorithm]{algorithm}

\usepackage{multirow}
\usepackage{multicol}

\usepackage{ulem}
\usepackage{color}

\algnewcommand{\IIf}[1]{\State\algorithmicif\ #1\ \algorithmicthen}
\algnewcommand{\EndIIf}{\unskip\ \algorithmicend\ \algorithmicif}

\usepackage{color, bm}
\definecolor{cadetblue}{rgb}{0.37, 0.62, 0.63}
\definecolor{burntorange}{rgb}{0.8, 0.33, 0.0}
\definecolor{darkchestnut}{rgb}{0.6, 0.41, 0.38}
\definecolor{darkcyan}{rgb}{0.0, 0.55, 0.55}
\definecolor{darkgoldenrod}{rgb}{0.72, 0.53, 0.04}
\definecolor{darklavender}{rgb}{0.45, 0.31, 0.59}
\definecolor{pansypurple}{rgb}{0.47, 0.09, 0.29}
\definecolor{tuscanred}{rgb}{0.51, 0.21, 0.21}

\newcounter{propcounter}
\refstepcounter{propcounter}

\newcommand{\given}{\,|\,}

\papertype{Original Article}
\paperfield{Journal Section}

\title{Maximum likelihood estimation in the additive hazards model}

\abbrevs{ML, maximum likelihood; MLE, maximum likelihood estimate; OLS, ordinary least squares}

\author[1]{Chengyuan Lu}
\author[1]{Jelle Goeman}
\author[1]{Hein Putter}

\affil[1]{Department of Biomedical Data Sciences, Leiden University Medical Center, Leiden, The Netherlands}

\corraddress{Chengyuan Lu, Department of Biomedical Data Sciences, Leiden University Medical Center, Leiden, The Netherlands}
\corremail{C.Lu@lumc.nl}

\runningauthor{Lu et al.}

\begin{document}
\maketitle

\begin{abstract}
The additive hazards model specifies the effect of covariates on the hazard in an additive way, in contrast to the popular Cox model, in which it is multiplicative. As non-parametric model, additive hazards offer a very flexible way of modeling time-varying covariate effects. It is most commonly estimated by ordinary least squares. In this paper we consider the case where covariates are bounded, and derive the maximum likelihood estimator under the constraint that the hazard is non-negative for all covariate values in their domain. We show that the maximum likelihood estimator may be obtained by separately maximising the log-likelihood contribution of each event time point, and we show that the maximising problem is equivalent to fitting a series of Poisson regression models with an identity link under non-negativity constraints. We derive an analytic solution to  the maximum likelihood estimator. We contrast the maximum likelihood estimator with the ordinary least squares estimator in a simulation study, and apply it to data on patients with carcinoma of the oropharynx.

\end{abstract}

\section{Introduction}

The fundamental concept in survival analysis is the hazard rate. There are several regression models describing the hazard rate, such as multiplicative risk models and additive risk models. The dominant hazard model in survival analysis is the multiplicative proportional hazards model~\citep{cox1972regression}. In many cases, however, the proportional hazards assumption is not met. In such cases, developing models that adequately describe the non-proportional effect of the covariate(s) is not straightforward \citep{gore1984regression, schemper1992cox, perperoglou2006reduced, vHP12} and different models should be considered. 

In this article, we focus on the non-parametric additive hazards model proposed by~\cite{Aalen1980, Aalen1989}, extensively studied in~\citet{martinussen2007dynamic}. The additive model defines the hazard rate as a linear form of the vector of covariates. Unconventionally, it does not naturally force the hazard rate to be positive. However, it has several useful properties, as set out by~\citet{aalen2008survival}. Firstly, the additive hazards model is internally consistent: it retains its additive structure if covariates are measured with uncertainty or are dropped from the linear expression. This contrasts with the proportional hazards model, which loses its proportional hazards property when covariates are omitted~\citep{strutherskalbfleisch1986, schumacher1987, bretagnolle1988}. This drawback of proportional hazards has triggered a recent debate about the danger of using the hazard ratio as a causal effect measure in survival analysis~\citep{hernan2010hazards, aalen2015does}. A second advantage of additive hazards is that it allows implemention of dynamic structures~\citep{martinussen2000nonparametric}, such as self-exciting processes, which are impossible to incorporate in most other nonlinear regression models.

The usual way to estimate the parameters in the additive hazards model is by Aalen's method, which uses ordinary least squares to estimate the cumulative effect of the covariates. Aalen's ordinary least squares method is straightforward to implement, but it has some disadvantages. The most important disadvantage is that it does not guarantee that the model-based hazards are positive for all time points for all subjects in the data, but only for the subjects that experience an event, and only at their event times. Furthermore, the ordinary least squares estimator is defined as a method-of-moments estimator and does not have the support of likelihood theory.

One method that generally yields efficient estimators is maximum likelihood. To the best of our knowledge, the maximum likelihood method has not been considered in any detail in the context of the additive hazards model, since no analytical expression for the maximum likelihood estimator was available. The objective of this paper is to derive the maximum likelihood estimator for the additive hazards model and to determine its advantages and limitations and in comparison to Aalen's ordinary least squares method. We find that there is no maximum solution to the survival likelihood in the unconstrained parameter space. However, we may find the solution if we choose a proper domain. A natural choice for such a domain is one that guarantees positivity of the hazard rate at all time points for all subjects in the data.

We show that the maximum likelihood estimator of the cumulative baseline hazard and cumulative covariate effects is a step function changing values only at the event time points, and that the maximum of the log-likelihood can be found by separately maximising the log-likelihood contributions corresponding to each of the event time points, as with Aalen's ordinary least squares solution. We define the constraint domain, and show that the maximisation problem is equivalent to fitting a series of identity-link Poisson regression models under non-negativity constraints. This problem has been studied before by~\citet{Marschner2010} and \citet{MARSCHNER2012}, who proposed an EM algorithm to obtain maximum likelihood estimators. In contrast, our solution is analytical, and computation time at each time point is linear in the sample size and in the number of covariates per time point. We also discuss the new method's connections to the ordinary least squares estimator. In particular, when only a single binary covariate is considered, we will show that maximum likelihood and ordinary least squares are equivalent. In Section~\ref{sec:simulation} we report on the results of a simulation study comparing maximum likelihood with ordinary least squares. Section~\ref{sec:application} illustrates our methods on data from a randomised clinical trial on patients with carcinoma of the oropharynx.

\section{Notation and model definition}
\label{sub:notation}

We use bold letters to indicate vectors and matrices. Define $T^*$ to be the time to event, $C$ to be the time to censoring, and let $T=\min(T^*, C)$, and $\Delta = I(T^* \leq C)$ the event indicator. We observe $(t_i, \delta_i, \mathbf{x}_i^*)$, $i=1,\ldots,n$, with $\mathbf{x}_i^* = (x_{i1},\ldots,x_{ip})^\top$ a $p$-vector of covariates. We assume that the covariates $x_{ij}$ are restricted to the interval $[0, 1]$. This looks like a severe restriction, but it is not, because for any covariate with a minimum of $a$ and a maximum of $b$ in the observed data, we can rescale the covariate to be $(x-a)/(b-a)$, which takes values in $[0, 1]$. Extend $\mathbf{x}_i^*$ with the constant $x_{i0} = 1$, obtaining $\mathbf{x}_i = (1, x_{i1},\ldots,x_{ip})^\top$. Define $\mathcal{E}$ to be the set of event time points $t_i$, with corresponding $\delta_i=1$ and let $|\mathcal E| = K $. We assume there are no ties, and define the ordered sequence of event time points $t_1^* < \ldots < t_K^*$. Finally, define $Y_i(t) =\sum_{i=1}^n I(t_i\geq t)$
as the at risk indicator of subject $i$, taking the value $1$ if subject $i$ is at risk for the event of interest just before time $t$ and the value $0$ otherwise.

The additive hazards model~\citep{Aalen1980, Aalen1989} assumes the hazard rate to be of the form
\[
    h(t \given \mathbf{x}_i^*) = \mathbf x^\top_i \bm \beta(t) =  \beta_0(t) + \beta_1(t)x_{i1} + \cdots + \beta_p(t)x_{ip}.
\]
The parameters $\beta_j(t)$ allow effects of the covariates to change over time, thus,  the additive  model is fully non-parametric. 

\section{Aalen's method}
\label{sub:OLS}

To estimate the parameters $\bm{\beta}(t)$, the most commonly used approach is Aalen's ordinary least squares method. Let us define the counting process $N_i(t)$ as the number of events experienced by subject $i$ before or at time $t$. Then the intensity $\lambda_i(t)$ of the counting process has the form
\[
        \lambda_i(t) =  Y_i(t)(\beta_0(t) + \beta_1(t)x_{i1} + \cdots + \beta_p(t)x_{ip})
            = Y_i(t) h(t_i \given \mathbf{x}_i^*).
\]
The formula $dN_i(t) = \lambda_i(t)dt + dM_i(t)$ gives equations
\[
    d\mathbf{N}(t) = \mathbf{X}(t)d\mathbf{B}(t) + d\mathbf{M}(t),
\]
where $\mathbf{N}(t) = (N_1(t),N_2(t),\ldots,N_n(t))^\top$ is the vector of counting processes, $\mathbf{X}(t)$ is the matrix of covariates multiplied with $Y_i(t)$ at each $i$th row, $\mathbf{B}(t) = (B_0(t),B_1(t),\ldots,B_p(t))^\top$ is the cumulative beta defined as $B_j(t) = \int_0^t \beta_j(s)ds$ and $\mathbf{M}(t) = (M_1(t), \ldots, M_n(t))^\top$, where $M_i(t)$ is the $i$th martingale error term. Ordinary least square regression thus gives Aalen's estimator of the $\bm{\beta}(t)$ if $X(t)$ is of full rank:
\[
    d\widehat{\mathbf{B}}(t) = (\mathbf{X}(t)^\top\mathbf{X}(t))^{-1} \mathbf{X}(t)^\top d\mathbf{N}(t).
\]
When $X(t)$ is not of full rank, we set $d\widehat{\mathbf{B}}(t) = 0$~\citep{aalen2008survival}.

\section{Maximum likelihood estimation}
\label{sub:ML}

In this section, we will derive an analytic solution for the maximum likelihood estimator under the natural constraint that the hazard is non-negative at each time point for all covariate values within the domain.

\subsection{The shape of the likelihood}

To begin with, let us consider the likelihood function for the additive hazards model. The log likelihood contribution of subject $i$ with an event or censored at time $t_i$ is given by
\[
    \ell_i = \delta_i \log\{h(t_i \given \mathbf{x}_i^*)\}+ \log\{ S(t_i \given \mathbf{x}_i^*)\},
\]
where $S(t \given \mathbf{x}_i^*) = P(T^* > t \given \mathbf{x}_i^*) = \exp(- H(t \given \mathbf{x}_i^*))$ is the survival probability and $H(t \given \mathbf{x}_i^*)$ the cumulative hazard of subject $i$. The log-likelihood is therefore given by~\citep{aalen2008survival}

\begin{equation}
    \label{eq:logliki}
     \ell = \sum_{i=1}^n \{\delta_i \log \mathbf \mathbf{x}_i^\top \bm{\beta}(t_i) - \mathbf{x}_i^\top \mathbf B(t_i)\}.
\end{equation}

Our first result is
\begin{proposition}
The likelihood function $\ell$ is unbounded.
\end{proposition}

\noindent%
We therefore introduce the natural constraint that the hazard of each possible covariate value of our data is non-negative at each time point, that is, we require that for all $t \geq 0$,
\begin{equation}
    \label{eq:hpos}
 h(t \given \mathbf{x}^*) \geq 0,\ \textrm{for}\ \mathbf{x}^* \in \{0, 1\}^p,
\end{equation}
which implies positivity of the hazard for $\mathbf{x}^* \in [0, 1]^p$. Our objective is to maximise the total log-likelihood given in~\eqref{eq:logliki}, subject to the constraint~\eqref{eq:hpos}.

Let us assume $\hat{\mathbf{B}}(t)$ is the function which maximises the likelihood function~\eqref{eq:logliki}, subject to the constraint~\eqref{eq:hpos}. Since $\hat{\mathbf{B}}(t)$ is an estimator of cumulative beta, which is generally the negative logarithm of the survival function, we may assume that it is right continuous with left limits (cadlag). Thus, we may decompose $\hat{\mathbf{B}}(t)$ as the summation of a continuous function $\hat{\mathbf{B}}_c(t)$ and a step function $\hat{\mathbf{B}}_s(t)$.

\begin{lemma}
\label{prop:beta0}
Under the constraint $h(t \given \mathbf{x}^*) \geq 0$ for $\mathbf{x}^* \in \{0, 1\}^p$, the log-likelihood can only achieve a maximum if $\hat{\mathbf{B}}_c(t) \equiv 0$ and $\hat{\mathbf{B}}_s(t)$ is a step function with jumps only at the event time points in $\mathcal{E}$.
\end{lemma}

\noindent%
Lemma~\ref{prop:beta0} implies that the maximisation problem is the same as maximising the total log-likelihood with respect to the jumps of $\mathbf{B}(t)$ at the event time points. Denoting the jump of $B_j(t)$ at the $k^{\textrm{th}}$ event time point $t_k^*$ as $\beta_{kj}$, $k=1,\ldots,K$, $j=0,\ldots,p$, this leads to $B_j(t) = \sum_{k: t_k^* \leq t} \beta_{kj}$.

The total log likelihood $\ell$ from~\eqref{eq:logliki} can now be rewritten in terms of the $\beta_{kj}$ as
\[
    \ell = \sum_{i=1}^n \Biggl\{ \delta_i \log \Bigl(\sum_{j=0}^p x_{ij} \beta_{k(i),j}\Bigr) - \sum_{j=0}^p x_{ij} \sum_{k: t_k^* \leq t_i} \beta_{kj} \Biggr\},
\]
where $k(i)$ is the index of the event time points $t_k^*$ corresponding to $t_i$, i.e.~$t_{k(i)}^* = t_i$ (in case $\delta_i=0$, $k(i)$ can be chosen as an arbitrary index in $\{1,\ldots,n\}$). We reorder this sum over subjects as a sum over the distinct event time points $t_k^*$, obtaining
\[
    \ell = \sum_{k=1}^{K} \Biggl\{ \log \Bigl(\sum_{j=0}^p x_{i(k),j} \beta_{kj}\Bigr) - \sum_{j=0}^p \beta_{kj} \sum_{l: t_l \geq t_k^*} x_{lj} \Biggr\} = \sum_{k=1}^{K} \ell_k^*,
\]
with
\begin{equation}\label{eq:ellk}
    \ell_k^* = \log \Bigl(\sum_{j=0}^p x_{i(k),j} \beta_{kj}\Bigr) - \sum_{j=0}^p \beta_{kj} s_{kj},
\end{equation}
$s_{kj} = \sum_{l: t_l \geq t_k^*} x_{lj}$,
and where $i(k)$ is the subject index corresponding to the $k^{\textrm{th}}$ event time point, i.e.~$t_{i(k)} = t_k^*$.

It is easy to see that each term $\ell_k^*$ is a function only of the variables $\beta_{kj}$, $j = 0,\ldots,p$. Since the terms $\ell_k^*$ do not have parameters in common, we may reduce the problem by separately maximising each $\ell_k^*$ as a function of $\bm{\beta}_k = (\beta_{k0}, \beta_{k1}, \ldots, \beta_{kp})$.

For the remainder of this section, we will fix (any) one of the time points $t_k^*$. To simplify notation, we are going to suppress dependence on $k$, and consider maximisation of
\begin{equation}\label{eq:ell}
    \ell^* = \log \Bigl( \sum_{j=0}^p x_j \beta_{j} \Bigr) - \sum_{j=0}^p s_j \beta_j =
    \log \Bigl( \mathbf{x}^\top \bm{\beta} \Bigr) - \mathbf{s}^\top \bm{\beta}
\end{equation}
with respect to $\bm{\beta} = (\beta_0, \beta_1, \ldots, \beta_p)$, where $\mathbf{s} = (s_0, s_1, \ldots, s_p)^\top$, $x_j$ stands for the $j^{\textrm{th}}$ element of the covariate vector of the subject that failed at time $t_k^*$, and $x_0=1$. 

\subsection{Maximum likelihood with the full constraint matrix}

\noindent%
From now on, we will give an alternative description of the constraint~(\ref{eq:hpos}) using a matrix form.  Let $\bm{\beta} = (\beta_0, \beta_1, \ldots, \beta_p)^\top$ be the vector of parameters of the function $\ell^*$. We can restate the constraint~(\ref{eq:hpos})   by the matrix $M_\mathbf{D}$ through the inequalities $M_\mathbf{D} \bm{\beta} \geq \bm 0$. We construct the matrix $M_\mathbf{D}$ such that the rows consist of all the $2^p$ possible $(p+1)$-tuples $(m_0, m_1, \ldots, m_p)$ in $\{0,1\}^{p+1}$ with $m_0 =1$.  So $M_\mathbf{D}$ is a $2^p \times (p+1)$ matrix. 

 The domain $\mathbf D\subset \mathbb R^{p+1}$ defined by constraint~\eqref{eq:hpos} is a polytope enclosed by the flat boundaries defined by $\gamma_i = M_i^\top \bm{\beta} = 0$, $i=1,\ldots,2^p$, where $M_i$ is the $i$th row of $M_{\mathbf D}$.  Now, we can show that the maximum point lies on the boundary of this polytope.

\begin{lemma} \label{prop:maximumpointReverse}
If the maximum point of $\ell^*$ in the domain $\mathbf D$ exists, then the maximum point  lies in the intersection of $p$ flat boundaries given by $\gamma_{i_j} = M_{i_j}^\top \bm \beta= 0$, $j = 1 \ldots, p$.
\end{lemma}

\noindent%
In Lemma~\ref{prop:maximumpointReverse}, we have given the geometric property of the maximum point of the $\ell^*$. The intersection of $p$ flat boundaries is a one-dimensional line. So, the maximum point always lies on a one-dimensional line edge on the boundary of the domain. The following lemma defines all the one-dimensional line edges on the boundary of the domain. There are $2p$ such line edges. 

\begin{lemma}\label{prop:geometryofedge}
The one-dimensional line edges on the boundary of the domain $M_\mathbf{D}\bm \beta \geq 0$ are given by $(0,\ldots,0,l,0,\ldots,0)$, with the position of $l$ ranging from $2$ to $p+1$, and $(l,\ldots,0,-l,0,\ldots,0)$, with the position of $-l$ ranging from $2$ to $p+1$. 
\end{lemma}

\noindent%
Let us denote by $\bm e_j$ the $(p+1)$-vector $(0, 0, \ldots, 0, 1, 0, \ldots, 0)^\top$, with the 1 at position $j+1$, and by $\bm f_j$ the $(p+1)$-vector $(1, 0, \ldots, 0, -1, 0, \ldots, 0)^\top$, with the -1 at position $j+1$, and define the sets $\mathcal{A}^+ = \{ \bm e_1, \ldots, \bm e_p \}$ and $\mathcal{A}^- = \{ \bm f_1, \ldots, \bm f_p \}$. We may call the set $\mathcal{A} = \mathcal{A}^+ \cup \mathcal{A}^-$, a finite set of cardinality $2p$, the set of admissible directions. Lemma's~\ref{prop:maximumpointReverse} and~\ref{prop:geometryofedge} imply that the maximum of $\ell^*$ lies on one of the edges ${\bm v} \cdot l$, with ${\bm v} \in \mathcal{A}$. We now come to our main theorem which describes the structure of the maximum likelihood estimator of the additive hazards model under the constraint of non-negativity of the hazards. Let $i(k)$ denote the index of the subject that fails at $t_k^*$, the maximum likelihood estimator $\hat{ \bm \beta}_k$ at the event time point $t_k^*$ can be determined as follows:

\begin{algorithm}[H]
\caption{Algorithm to find the maximum point within the domain $\mathbf D$.\label{alg:s1}}

\begin{itemize}
    \item Calculate $\mathbf s_{k} = \sum_{l: t_l \geq t_k^*} \mathbf x_{l}$; 
    \item Calculate the values of ratio  $\bm v_j = x_{i(k), j} / s_{k,j}$ for $1 \leq j \leq p$;
    \item Calculate the values of ratio  $\bm v_{p + j} = (x_{i(k), 0} - x_{i(k), j}) / (s_{k,0} - s_{k,j})$ for $1 \leq j \leq p$;
    \item Find the set $M$ of indices $m = 1, \ldots, 2p$, for which $\bm v_m$ is maximised;
    \item For any index $m \in M$, calculate the value of $\hat{\bm \beta}_{k}^{(m)}$ by 
\begin{equation*}\hat{\bm \beta}_{k}^{(m)} =  \begin{cases}
            \bm e_m / s_{k,m}, & \hbox{if $1 \leq m \leq p$;} \\
            \bm f_{m-p} / (s_{k,0} - s_{k,(m-p)}),& \hbox{if $p + 1 \leq m \leq 2p$;}
          \end{cases}
         \end{equation*}
    \item Return $\hat{\bm \beta}_{k}^{(m)}$, $m \in M$.
\end{itemize}

\end{algorithm}

\begin{theorem}\label{thm:maintheorem}
For an additive hazards model $h(t \given \mathbf{x}_i^*) = \bm \beta(t)\mathbf{x}_i$ with constraints  $h(t \given \mathbf{x}^*) \geq 0,\ \textrm{with }\ \mathbf{x}^* \in [0, 1]^p$, assuming no ties,  the values returned by Algorithm \ref{alg:s1} are maximum likelihood estimators. 
\end{theorem}

\noindent%
Theorem \ref{thm:maintheorem} offers some insight into the maximum likelihood estimator. There are only two possible patterns of the jump $\bm \beta(t_k^*)$. Either there is an increment of the coefficient $\beta_j(t_k^*)$ for one covariate or there is a decrease of $\beta_j(t_k^*)$ for one covariate which is compensated by an increment of the intercept. So, this estimator is sparse in the change of $\bm \beta(t)$.

Theorem \ref{thm:maintheorem} also implies a condition for the existence and uniqueness of the likelihood estimator. The likelihood estimator exists if and only if $s_{k,m} \neq 0$ and $s_{k,0}- s_{k,(m-p)} \neq 0$. which correspond to the cases all $x_i$'s are zeros or all $x_i$'s are ones, $i \geq i(k)$. Uniqueness is not guaranteed by Theorem \ref{thm:maintheorem}. In fact,  all choices of the ratios $x_{i(k), j} / s_{k,j}$ and/or $(x_{i(k), 0} - x_{i(k), j}) / (s_{k,0} - s_{k,j})$ found by Algorithm \ref{alg:s1} result in the same maximum. In that case any ${\bm v} \in \mathcal{A}$ corresponding to such a maximum is allowed, as well as any convex combination, as stated in the following theorem.

\begin{theorem}\label{thm:mutisolution}
If $\hat{\bm\beta}_{k}^{(m_1)}$ and $\hat{\bm\beta}_{k}^{(m_2)}$ are solutions given by Algorithm \ref{alg:s1} which maximise the likelihood $\ell$, then any convex combination $\lambda \bm \hat{\bm\beta}_{k}^{(m_1)} + (1 - \lambda) \hat{\bm\beta}_{k}^{(m_2)}$, with $0 \leq \lambda \leq 1$, attains the same maximised likelihood.
\end{theorem}

In practice, we suggest an average of all solutions given by Theorem~\ref{thm:maintheorem}. By Theorem~\ref{thm:mutisolution}, this is also a solution which maximises the likelihood.

With our result on the maximum likelihood estimator, we also have the following result which reveals the connection between the maximum likelihood estimator and Aalen's estimator.

\begin{proposition}
For an additive hazards model with one binary covariate, the maximum likelihood estimator and the Aalen estimator coincide.
\end{proposition}

This can also be explained intuitively by Theorem \ref{thm:maintheorem} as follows. For one binary covariate, Aalen's ordinary least squares method can be viewed as selecting one of two competing trends at each event time. Theorem~\ref{thm:maintheorem} shows that the nature of maximum likelihood estimator is similar; the jump of the coefficient takes place for only one covariate, i.e., only one trend is selected at each event time.

\subsection{Example}

We illustrate Theorem~\ref{thm:maintheorem} using a simple example with two covariates. Let us assume that at a certain event time point $t^*$ a subject fails with covariate values $x_1=0$ and $x_2=1$, so we have $\mathbf{x}^* = (0,1)$, and $\mathbf{x} = (1,0,1)$. Suppose that $\mathbf{s} = (8,5,6)^\top$. Then the log-likelihood is given by 
\begin{equation}\label{eq:twocovml}
    \ell^* = \log (\beta_0 + \beta_2)- 8 \beta_0 -5\beta_1 -6 \beta_2.
\end{equation}

In this case, the domain is defined by the matrix $\mathbf{M_D}$ through the boundary conditions $M_{\mathbf D} \bm \beta \geq 0$, where 
\[ M_{\mathbf D} = \begin{pmatrix}
    1 & 0 & 0 \\
    1 & 0 & 1 \\
    1 & 1 & 0 \\
    1 & 1 & 1 \\
\end{pmatrix}.\]

Theorem~\ref{thm:maintheorem} directs us to compare the values of $x_1/s_1$, $x_2/s_2$, $(x_0-x_1)/(s_0-s_1)$, and $(x_0-x_2)/(s_0-s_2)$, namely $0$, $\frac{1}{6}$, $\frac{1}{3}$, $0$, corresponding to the one-dimensional line edges ${\bm v} \cdot l$, with $\bm v = (0,1,0)^\top, (0,0,1)^\top, (1,-1,0)^\top$ and $(1,0,-1)^\top$, respectively, in the set of admissible directions $\mathcal{A}$. The third of these numbers is largest, so the one-dimensional line ${\bm v} \cdot l$, with $\bm v = (1,-1,0)^\top$ is the edge where the maximum point lies. Maximising over $l$ gives $l = \frac{1}{\mathbf{s}^{\top} \bm{v}} = \frac{1}{3}$, $\hat{\bm \beta}(t) = (\frac{1}{3}, -\frac{1}{3},0)$ and the maximised log-likelihood $\ell^* = \log(\frac{1}{3})- 1$.

\subsection{Maximum likelihood with a partial constraint matrix}
\label{sub:MLEfind} 

\noindent%
In previous sections, we show that the maximum point of the likelihood function lies in one of the $2p$ one-dimensional line edges on the boundary. This result is derived for the case where the domain is given by the matrix $M_{\mathbf D}$ whose rows are all the combinations of $1$ and $0$ in the last $p$ columns. Sometimes we need more flexible domains, for example when dummy covariates are introduced or when it is known that some coefficients are positive. In this more general case Theorem \ref{thm:maintheorem} does not apply, and numerical solutions are needed. 

We consider a general domain defined by $M_D \bm \beta \geq 0$ for some matrix $M_D$, where the matrix $M_D$ can be an arbitrary matrix of $p+1 $ columns. We assume that the rank of $M_D$ is $p+1$. We assert that Lemma \ref{prop:maximumpointReverse} still applies for this constraint matrix $M_D$. The maximum point lies on a one-dimensional edge on the boundary, defined as the intersection of $p$ flat boundaries. So, any local maximum point of a   one-dimensional edge is a potential solution of the maximum likelihood of the domain $D$.  Given any one-dimensional edge, we have $p$ equations $\gamma_i = M_{i} \bm \beta = 0$. Combined with the first order condition which is also a linear equation, we have $p+1$ linear equations whose solution is the maximum point on this edge. By testing its derivatives and constraint conditions, we can check if it is a maximum point of the domain $D$.  However, it is difficult to find the right edge on which the maximum point lies. We introduce three methods to find the right edge: the na\"{\i}ve method, the ascending method and the descending method. 

The na\"{\i}ve method loops through all the possible combinations of $p$ rows of $M_D$. Every combination of $p$ rows defines a one-dimensional edge and gives a possible solution of the maximum point $\bm \beta^*$. We can test its derivatives and constraint conditions to check if it is the maximum point we want to find. This na\"{\i}ve method is easy to understand, but not efficient, since the complexity grows approximately with the $p$th power of number of the rows of $M_D$. 

sThe ascending method searches for a series of $\bm \beta^*$ with increasing likelihood. More specifically, the algorithm  starts from a $p$-combination of rows, whose maximum point is $\bm \beta^*_0$. We assume this edge is inside the domain. If  $\bm \beta^*_0$ is not the maximum point of the domain $D$, its gradient points to the inside of the domain. Following this direction, we can find an adjacent one-dimensional edge whose maximum point $\bm \beta^*_1$ has a larger likelihood.  Since this algorithm starts from an edge already on the domain, as a result, it is quite efficient when a good starting edge of the domain is known. To find a new edge, we use the gradient of the log likelihood, constrained to the adjacent hyperplanes. This method works well if the geometry of the domain is not too complicated. 

The descending method is the reverse of the ascending method. Assuming that we find a maximum point $\bm \beta^*$ on a one-dimensional edge, it is not necessary that this $\bm \beta^*$ satisfies all the constraint conditions $M_D \bm \beta^* \geq 0$. However, it may satisfy some of the constraint conditions, i.e., we have $\gamma_i = M_i \bm \beta^* \geq 0$ for some row vector $M_i$. So, for a submatrix $M_D'$ consisting of these  rows, we have $M_D' \bm \beta \geq 0$. In many cases, this means we find a maximum point of the domain $D'$ defined by $M_D' \bm \beta \geq 0$. This new domain $D'$ is larger than the domain $ D$ and we have  $D' \supset D$. So the maximum likelihood of domain $D'$ is larger than the maximum likelihood of domain $D$. Starting from this domain $D'$, we can add the constraints step by step, which gives a smaller and smaller domains and finally reaches the domain $D$ that we want. Correspondingly, we have a series of decreasing maximum likelihood. This gives the name of descending method. The descending method also requires a starting edge whose maximum point is the maximum point of some larger domain $D' \supset D$. We can apply the na\"{\i}ve method to find such an edge and this is the time consuming part of the algorithm.

Details and pseudo-code of all three methods are given in the Supplemental Information.

\section{Simulation}
\label{sec:simulation}

The purpose of the simulation study is to show feasibility of our maximum likelihood estimator and compare its performance with the ordinary least squares estimator. We assume $\bm \beta(t) = (0.05,0.02,0.04,0.06,0.08)t$, hence the true hazard $h$ is given by:
\[
    h(t \,|\, \mathbf{x}) = \sum_i \beta_i x_i t= (0.05 + 0.02\  x_1 + 0.04\  x_2 + 0.06 \  x_3 +0.08\  x_4)t.
\]
This corresponds to a Weibull distribution with shape $b=2$ and rate $a = (0.05 + 0.02 x_1 + 0.04 x_2 + 0.06 x_3 + 0.08 x_4) / 2$, using the parametrisation $h(t; a, b) = a b t^{b-1}$ for the Weibull hazard.

We randomly generate covariate data $\mathbf{x}_i$, for $i=1,\ldots, n$, with $n=500$, with independent uniform distributions on $[0, 1]$. For each subject an independent censoring time $C_i$ is generated with uniform distribution over $(2.5,7.5)$, giving a censoring rate of about $22\%$. For Aalen's ordinary least squares method we use the implemention in the \textbf{timereg} package~\citep{martinussen2007dynamic}. We use the maximum likelihood estimator with full constraint matrix as introduced in Section~\ref{sub:ML}.

\begin{figure}[!ht]
\centering
\includegraphics[width=0.95\textwidth]{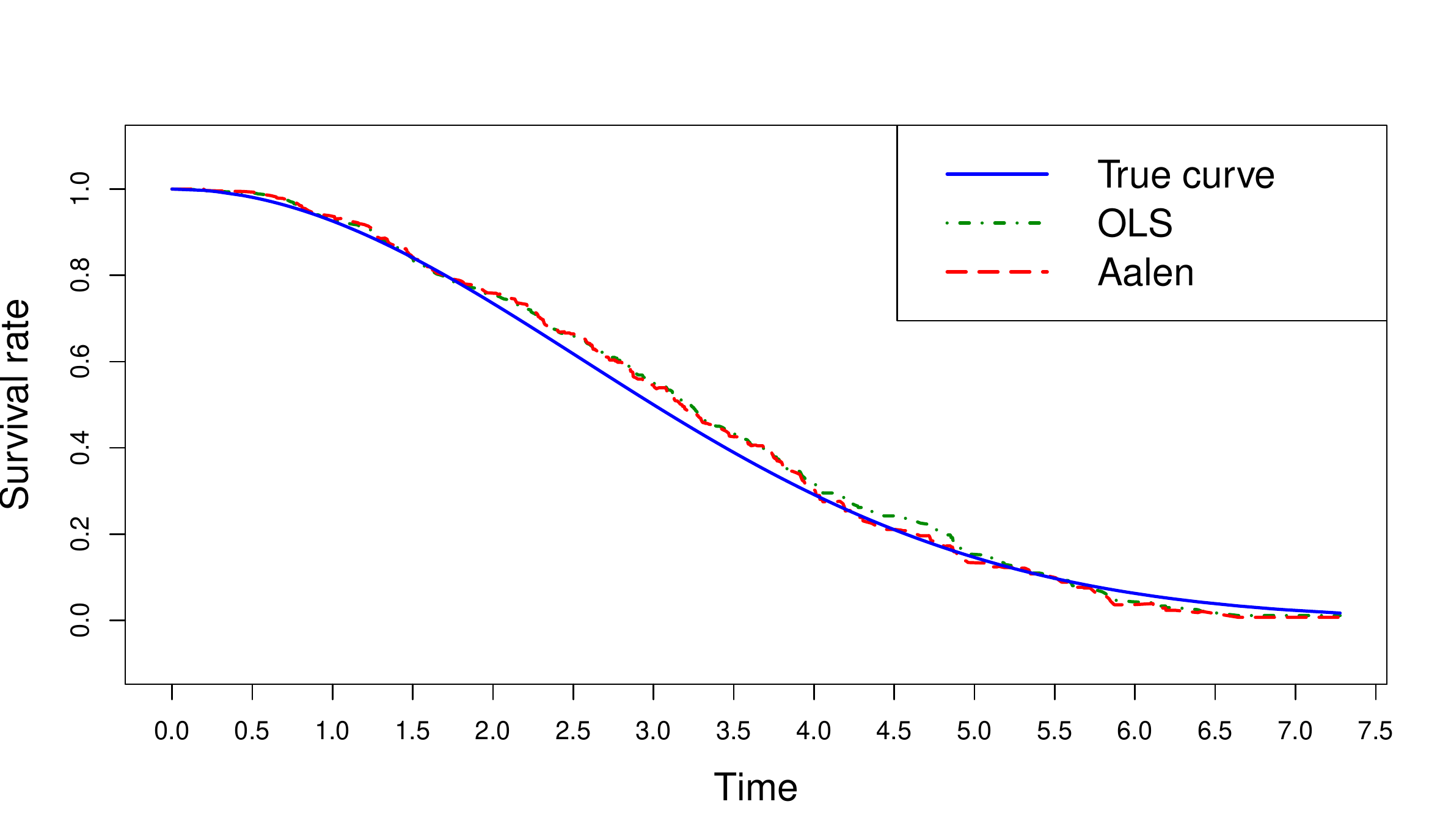}
\caption{Estimated survival curves by ordinary least squares (OLS) and by maximum likelihood (MLE)}
\label{fig:expsurCCM}
\end{figure}

We choose a subject with data of values $x = (0.4,0.6,0.4,0.6)$. 
Figure~\ref{fig:expsurCCM} shows the true survival curve and a single realisation of the estimated survival curves by the two methods. The survival curves given by both estimators are similar and approximate the true survival curve well.

\begin{figure}[H]
\centering
\includegraphics[width=0.95\textwidth]{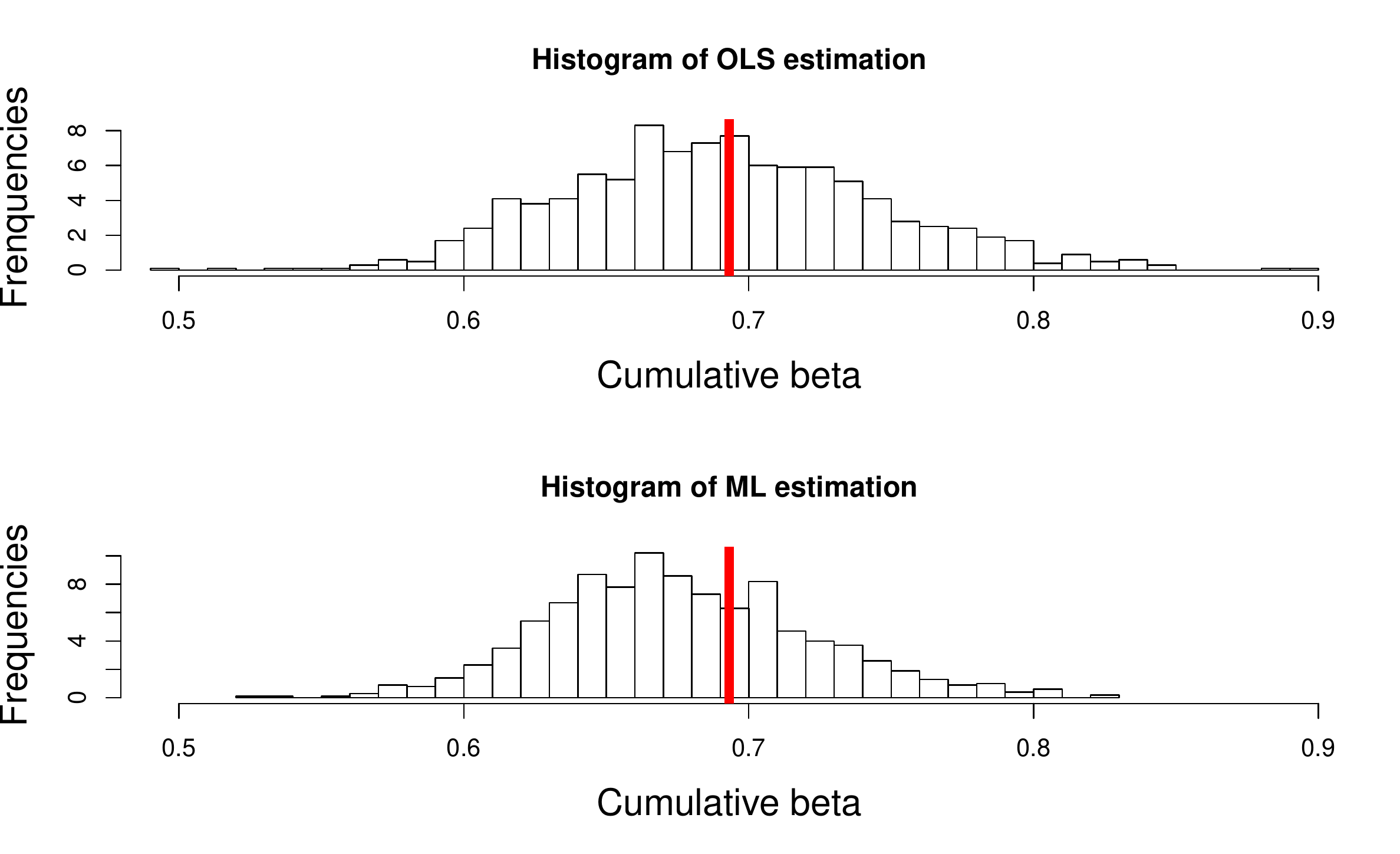}
\caption{Histograms of estimated cumulative hazards at time $t=3$ (median) by ordinary least squares (OLS) and maximum likelihood (ML).}\label{fig:hist}
\end{figure}

We repeat the same simulation set-up for $1000$ replications with the same subject. We check the estimated cumulative hazards at the three quartiles of the true survival function for that subject, at $t=1.93$, $3.00$ and $4.24$. Figure~\ref{fig:hist} shows histograms of the estimated cumulative hazards for the two methods at the median. The true value of the cumulative hazard is $0.693$, shown by the red vertical line. The ordinary least squares estimate is quite symmetric around the true value, while the maximum likelihood estimator is slightly skewed to the left. However, the variability of the maximum likelihood estimation is smaller.

This behaviour is also seen in Table~\ref{tab:cumulativebeta}, which shows that for all three time points, the maximum likelihood estimator has more bias but smaller variance. As a result, the root mean squared error is about 10 to 16 percent smaller for the maximum likelihood estimator, compared to ordinary least squares.
\begin{center}
\begin{tabular}{p{1.5cm}<{\centering} c p{2cm}<{\centering} c p{2cm}<{\centering} c c c  } 
\hline
True survival probability & Time point & True cumulative hazard & Method & Estimated cumulative hazard & Bias & Empirical SE & RMSE\\ 
\hline
\multirow{2}*{$0.75$}    & \multirow{2}*{$1.93$} & \multirow{2}*{$0.288$} & OLS & $0.286$ & $-0.002$ & $0.031$ & $0.031$\\
             &     &  & MLE & $0.280$  & $- 0.007$ & $0.025$ & $0.026$\\
\hline
  \multirow{2}*{$0.5$}    & \multirow{2}*{$3.00$} & \multirow{2}*{$0.693$} & OLS & $0.691$ & $-0.002$ & $0.056$ & $0.056$\\
             &     &  & MLE & $0.676$  & $- 0.017$ & $0.046$ & $0.049$\\
\hline
   \multirow{2}*{$0.25$}    & \multirow{2}*{$4.24$} & \multirow{2}*{$1.386$} & OLS & $1.384$ & $-0.002$ & $0.107$ & $0.107$\\
             &     &  & MLE & $1.353$  & $- 0.034$ & $0.089$ & $0.095$\\
\hline
\end{tabular}
\captionof{table}{Bias (mean estimate minus true value), empirical SE (the standard deviation of the estimates around their mean), and RMSE (root mean squared error, square root of the mean squared difference between estimate and truth) of cumulative hazards estimators by ML and OLS.}
\label{tab:cumulativebeta}
\end{center}

\section{Application}
\label{sec:application}

We apply our methods to data from a clinical trial with 195 patients with carcinoma of the oropharynx by the Radiation Therapy Oncology Group in the United States. The data are introduced by~\citet[Section 1.1.2, Appendix A]{KalbfleishPrentice2012} and used there for illustration of additive hazards models in~\citet{aalen2008survival}. Patients were randomised into two treatment groups (“standard” and “experimental” treatment), and survival times were measured in days from diagnosis. Seven covariates were included in the data: Sex, Treatment, Grade, Age, Condition, T-stage and N-stage. Following~\citet{aalen2008survival}, we take all the covariates as continuous. We re-scale them to $[0,1]$. 

\begin{figure}[!ht]
\centering
\includegraphics[width=0.95\textwidth]{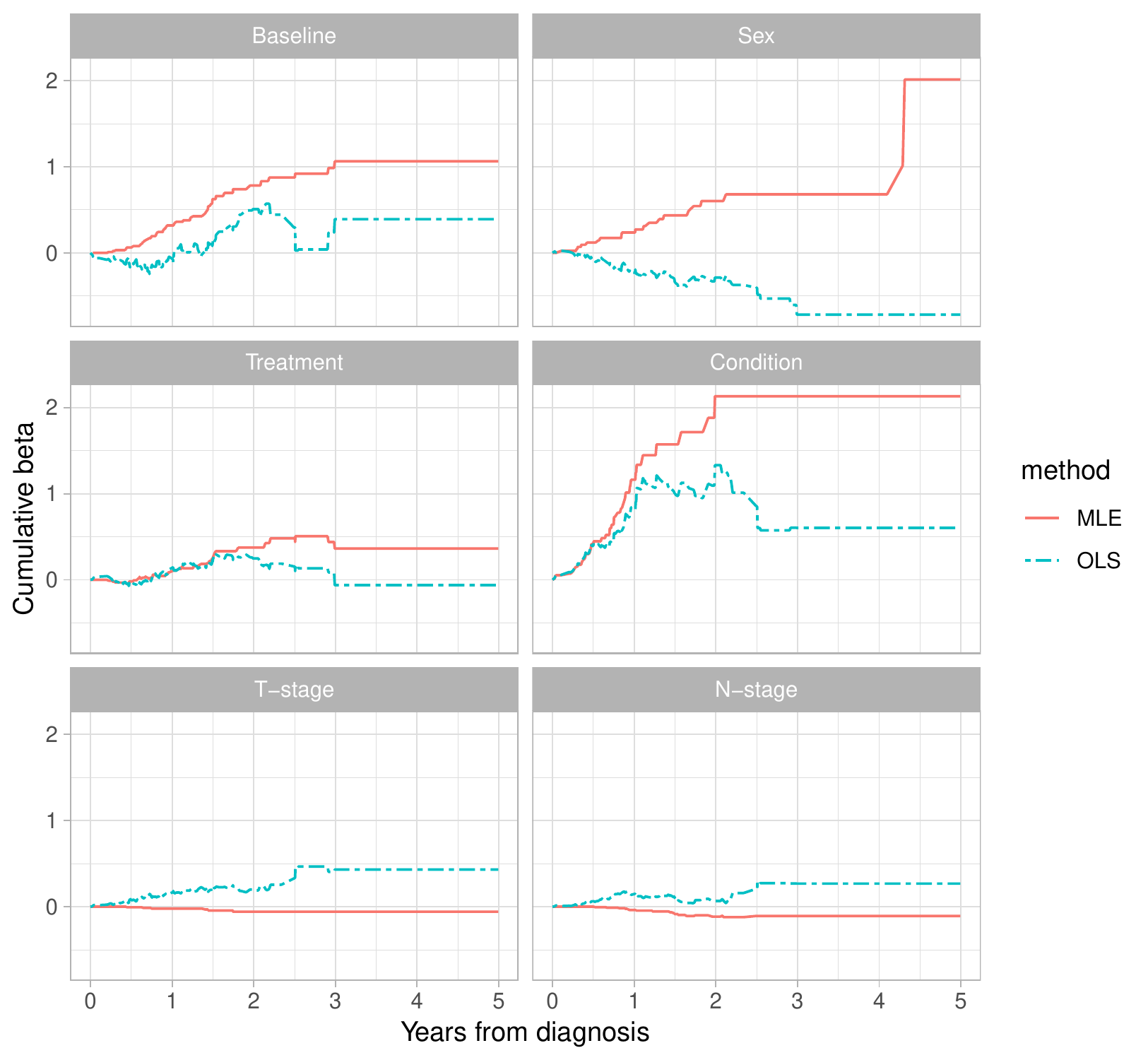}
\caption{Estimated cumulative beta's of factors by maximum likelihood (MLE) and ordinary least squares (OLS)}\label{fig:cbeta_orapha}
\end{figure}

The cumulative beta's of the baseline and covariates sex, treatment group, condition, T-stage and N-stage are shown in Figure~\ref{fig:cbeta_orapha}. The ordinary least squares baseline hazard estimate decreases over the first nine months and between 2 and 2.5 years, which would correspond to a negative hazard for subjects with all covariates 0, and an increasing survival curve. The maximum likelihood estimator of the baseline hazard is monotone increasing over time, as desired. For treatment effect, condition, T-stage and N-stage factors the two estimated curves of the cumulative beta are very close to each other in theses cases. For sex, the two methods show opposite trends. The differences in estimates between the two methods may partly be explained by the large number of covariates in the model, in relation to the modest sample size, leading to highly variable estimates for both methods.

\begin{figure}[H]
\centering
\includegraphics[width=0.95\textwidth]{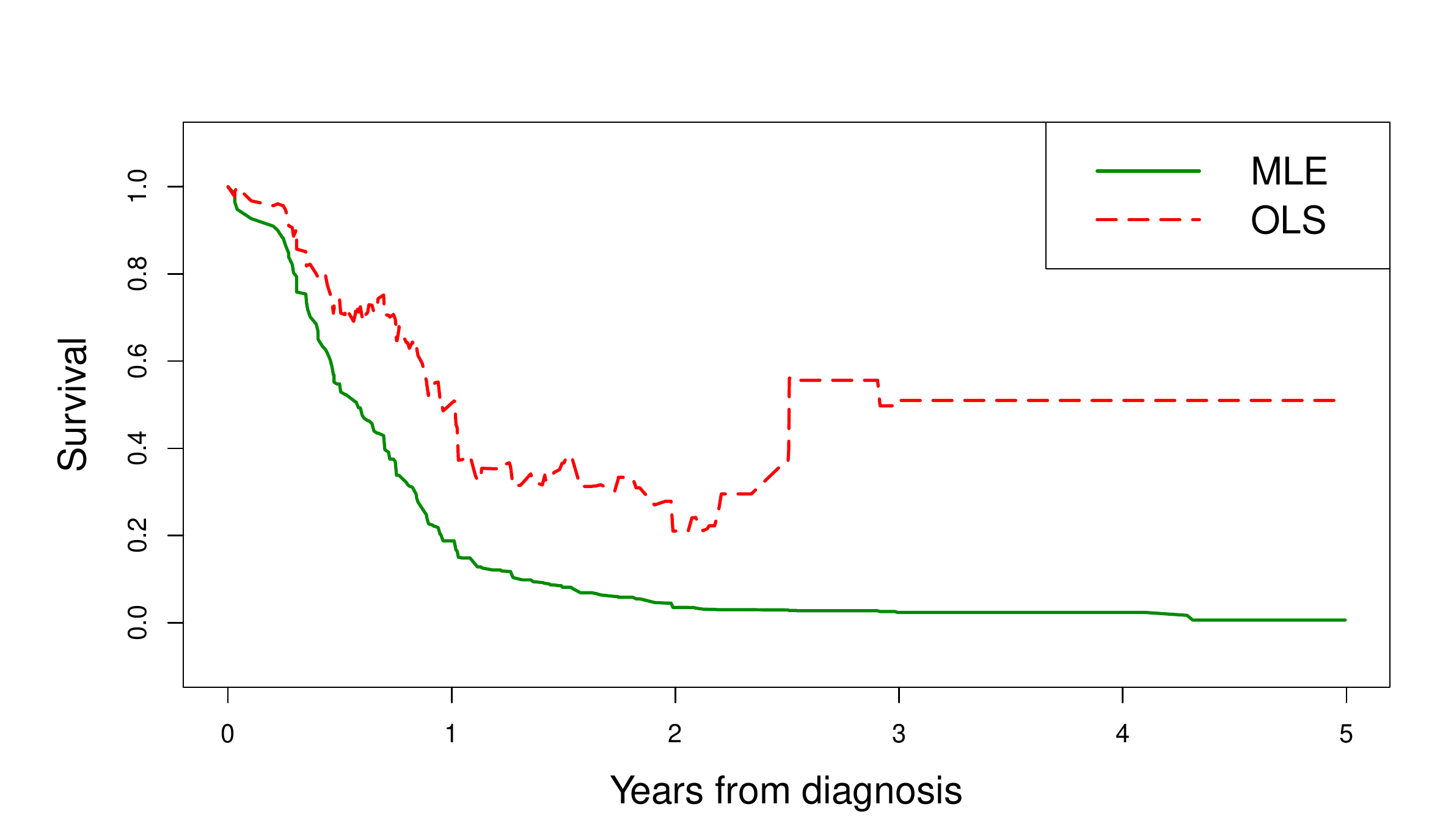}
\caption{Estimated survival curves by the two methods maximum likelihood (MLE) and ordinary least squares (OLS).}\label{fig:estsurv}
\end{figure}

To illustrate the difference in the behavior of the survival curves between maximum likelihood and ordinary least squares, we choose a subject who is female of age $55$ in the treatment group, of cancer grade moderate differentiated, of condition restricted worked, of second T-stage and first N-stage. The covariates of the subject are given by $\mathbf{x} = (1, 1, 0, 1, -0.5, 1, 1, 1)$. The estimated survival curves according to the ordinary least squares and maximum likelihood models are shown in Figure~\ref{fig:estsurv}. The two estimators give  similar results in the first half year when the risk set is large. After that, the ordinary least squares survival curve starts to fluctuate wildly, while the maximum likelihood estimate shows a steady decrease.

\section*{Acknowledgements}

The project is sponsored by the Leiden University Fund / Hakkenberg $\alpha$ $\beta$ $\gamma$-integratie Fonds, www.luf.nl. We would like to thank and acknowledge their support.\vspace*{-8pt}

\section*{conflict of interest}
No conflict of interest.



\bibliography{AdditiveHazard}

\begin{thebibliography}{18}
\expandafter\ifx\csname natexlab\endcsname\relax\def\natexlab#1{#1}\fi
\expandafter\ifx\csname url\endcsname\relax
  \def\url#1{\texttt{#1}}\fi
\expandafter\ifx\csname urlprefix\endcsname\relax\def\urlprefix{URL: }\fi

\bibitem[{Aalen(1980)}]{Aalen1980}
Aalen, O. (1980) A model for nonparametric regression analysis of counting
  processes.
\newblock In \textit{Mathematical Statistics and Probability Theory. Lecture
  Notes in Statistics} (eds. W.~Klonecki, A.~Kozek and J.~Rosiński), vol.~2, 1
  -- 25. Springer, New York.

\bibitem[{Aalen(1989)}]{Aalen1989}
--- (1989) A linear regression model for the analysis of life times.
\newblock \textit{Statistics in Medicine}, \textbf{8}, 907 -- 925.

\bibitem[{Aalen et~al.(2008)Aalen, Borgan and Gjessing}]{aalen2008survival}
Aalen, O., Borgan, O. and Gjessing, H. (2008) \textit{Survival and Event
  History Analysis: a Process Point of View}.
\newblock Springer, New York.

\bibitem[{Aalen et~al.(2015)Aalen, Cook and R{\o}ysland}]{aalen2015does}
Aalen, O.~O., Cook, R.~J. and R{\o}ysland, K. (2015) Does {C}ox analysis of a
  randomized survival study yield a causal treatment effect?
\newblock \textit{Lifetime Data Analysis}, \textbf{21}, 579--593.

\bibitem[{Bretagnolle and Huber-Carol(1988)}]{bretagnolle1988}
Bretagnolle, J. and Huber-Carol, C. (1988) Effects of omitting covariates in
  {C}ox's model for survival data.
\newblock \textit{Scandinavian Journal of Statistics}, \textbf{15}, 125--138.

\bibitem[{Cox(1972)}]{cox1972regression}
Cox, D.~R. (1972) Regression models and life-tables.
\newblock \textit{Journal of the Royal Statistical Society: Series B
  (Methodological)}, \textbf{34}, 187--202.

\bibitem[{Gore et~al.(1984)Gore, Pocock and Kerr}]{gore1984regression}
Gore, S.~M., Pocock, S.~J. and Kerr, G.~R. (1984) Regression models and
  non-proportional hazards in the analysis of breast cancer survival.
\newblock \textit{Journal of the Royal Statistical Society: Series C (Applied
  Statistics)}, \textbf{33}, 176--195.

\bibitem[{Hern{\'a}n(2010)}]{hernan2010hazards}
Hern{\'a}n, M.~A. (2010) The hazards of hazard ratios.
\newblock \textit{Epidemiology (Cambridge, Mass.)}, \textbf{21}, 13 -- 15.

\bibitem[{van Houwelingen and Putter(2012)}]{vHP12}
van Houwelingen, H. and Putter, H. (2012) \textit{Dynamic Prediction in
  Clinical Survival Analysis}.
\newblock CRC Press, Boca Raton.

\bibitem[{Kalbfleisch and Prentice(2012)}]{KalbfleishPrentice2012}
Kalbfleisch, J.~D. and Prentice, R.~L. (2012) \textit{The Statistical Analysis
  of Failure Time Data}.
\newblock Wiley, New York.

\bibitem[{Marschner(2010)}]{Marschner2010}
Marschner, I.~C. (2010) Stable computation of maximum likelihood estimates in
  identity link {P}oisson regression.
\newblock \textit{Journal of Computational and Graphical Statistics},
  \textbf{19}, 666--683.

\bibitem[{Marschner et~al.(2012)Marschner, Gillett and
  O’Connell}]{MARSCHNER2012}
Marschner, I.~C., Gillett, A.~C. and O’Connell, R.~L. (2012) Stratified
  additive {P}oisson models: Computational methods and applications in clinical
  epidemiology.
\newblock \textit{Computational Statistics \& Data Analysis}, \textbf{56}, 1115
  -- 1130.

\bibitem[{Martinussen and Scheike(2007)}]{martinussen2007dynamic}
Martinussen, T. and Scheike, T.~H. (2007) \textit{Dynamic {R}egression {M}odels
  for {S}urvival {D}ata}.
\newblock Springer, New York.

\bibitem[{Martinussen et~al.(2000)Martinussen, Scheike
  et~al.}]{martinussen2000nonparametric}
Martinussen, T., Scheike, T.~H. et~al. (2000) A nonparametric dynamic additive
  regression model for longitudinal data.
\newblock \textit{The Annals of Statistics}, \textbf{28}, 1000--1025.

\bibitem[{Perperoglou et~al.(2006)Perperoglou, Le~Cessie and van
  Houwelingen}]{perperoglou2006reduced}
Perperoglou, A., Le~Cessie, S. and van Houwelingen, H.~C. (2006) Reduced-rank
  hazard regression for modelling non-proportional hazards.
\newblock \textit{Statistics in Medicine}, \textbf{25}, 2831--2845.

\bibitem[{Schemper(1992)}]{schemper1992cox}
Schemper, M. (1992) Cox analysis of survival data with non-proportional hazard
  functions.
\newblock \textit{Journal of the Royal Statistical Society: Series D (The
  Statistician)}, \textbf{41}, 455--465.

\bibitem[{Schumacher et~al.(1987)Schumacher, Olschewski and
  Schmoor}]{schumacher1987}
Schumacher, M., Olschewski, M. and Schmoor, C. (1987) The impact of
  heterogeneity on the comparison of survival times.
\newblock \textit{Statistics in Medicine}, \textbf{6}, 773--784.

\bibitem[{Struthers and Kalbfleisch(1986)}]{strutherskalbfleisch1986}
Struthers, C.~A. and Kalbfleisch, J.~D. (1986) Misspecified proportional
  hazards models.
\newblock \textit{Biometrika}, \textbf{73}, 363--369.

\end{thebibliography}

\end{document}